\documentclass[aps,prl,twocolumn,superscriptaddress]{revtex4}
\usepackage{mathrsfs}
\usepackage{epsfig}
\usepackage{graphicx}
\usepackage{amsfonts}
\usepackage[figuresright]{rotating}
\usepackage{amssymb}
\usepackage{amsmath}
\usepackage{dcolumn}
\usepackage{bm}
\usepackage{color}
\usepackage{xcolor}

\usepackage[pdftex,colorlinks=true]{hyperref}

\def\be{\begin{equation}} \def\ee{\end{equation}}
\def\bea{\begin{eqnarray}} \def\eea{\end{eqnarray}}

\newcommand{\WQCASQC} { Wilczek Quantum Center and Key Laboratory of Artificial Structures and Quantum Control, School of Physics and Astronomy, Shanghai Jiao Tong University, Shanghai 200240, China}

\begin{document}
%\title{Imaginary time crystal: {\clb an exotic phase of quantum matter}}
%\title{Nonequilibrium magnets with emergent frustration}
\title{Memory-driven Topological Defects and Unconventional Long-Range Order}
\author{Ziyang Ding}
\affiliation{\WQCASQC}

\author{Zi Cai}
\email{zcai@sjtu.edu.cn}
\affiliation{\WQCASQC}

\begin{abstract}
We investigate many-body systems with time-delayed self-interactions mediated by a memory-feedback mechanism. We show that such temporal interactions generate non-equilibrium orders and unique topological defects absent in equilibrium—specifically, helical vortices wherein opposite vorticities propagate in reverse directions along domain walls. In one dimension, memory feedback stabilizes true long-range order against weak noise, thereby circumventing the Mermin-Wagner theorem and resulting in an unconventional finite-temperature phase transition with a dynamical exponent $z = 4$. Physical realizations of these memory-driven non-equilibrium systems using active mechatronic metamaterials has also been proposed. These results demonstrate engineered temporal interactions as a powerful paradigm for non-equilibrium many-body physics.

%Using spin-wave analysis, we elucidate the mechanisms driving the diverse ordered phases discovered in this model. Furthermore, we characterize a noise-induced phase transition featuring an unconventional critical behavior governed by a dynamical critical exponent of $z=4$. Finally, we discuss potential physical realizations of this memory-driven non-equilibrium system.

\end{abstract}

%\pacs{05.30.Jp, 75.10.Pq, 02.70.Ss, 03.65. Yz}

\maketitle

{\it Introduction--} Feedback mechanisms—where the evolution of a system is dynamically modulated by its past or present outputs—play a foundational role across modern science and engineering\cite{Giovannetti1999,Terhal2015,Magann2022,Yamaguchi2023,Wu2022,Zhao2023,Lee2008,Guan2016,Mijalkov2016,Bar2020,Kopp2023}. In practical realizations, an explicit or implicit time delay in the feedback channel is often inevitable, naturally rendering the underlying system dynamics non-Markovian through persistent history dependence. While the non-Markovian memory effects are usually considered as a computational burden or a source of unwanted decoherence to be mitigated via the Markovian approximation, recent developments demonstrate that persistent history-dependence can serve as a powerful resource to engineer entirely new non-equilibrium physics\cite{Grimsmo2015,Ivanov2020}. Despite this potential, the memory effect remains largely unexplored in the context of many-body physics, with only a few exceptions\cite{Ma2026,Sun2025,Sipling2025}. Whether such history-dependent non-equilibrium mechanisms can lead to physics beyond the scope of conventional equilibrium frameworks remains an open question.

\begin{figure}[htbp]
	\centering
	\includegraphics[width=0.49\textwidth]{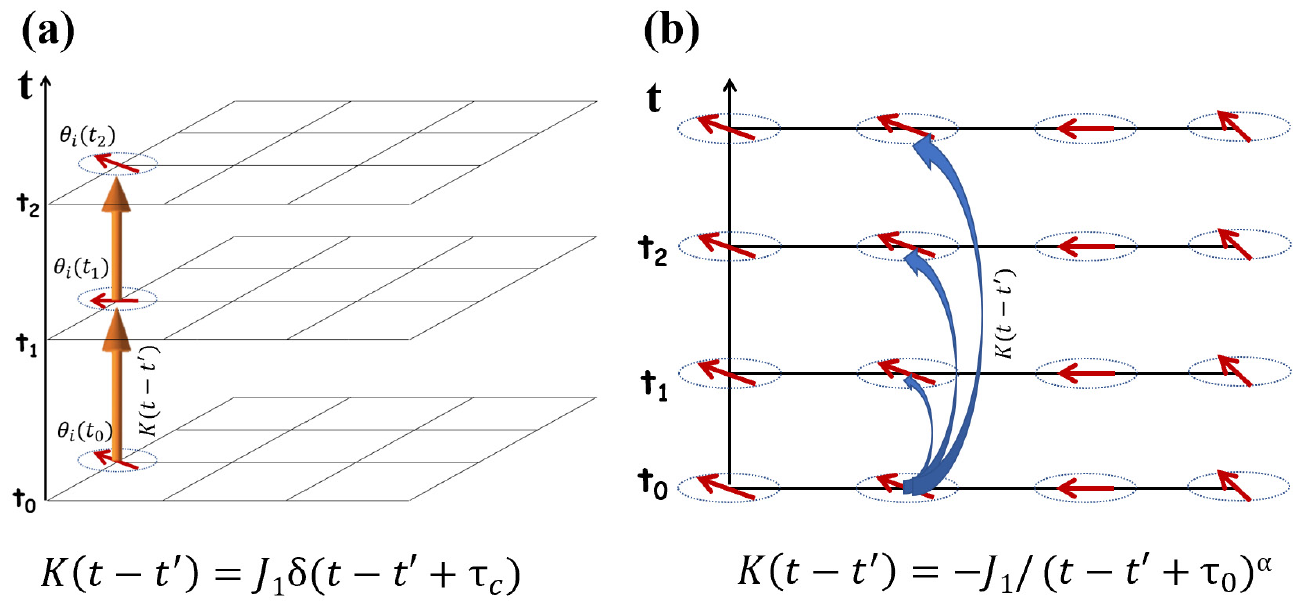}
	\caption{ Sketch of (a)  2D model with short-range antiferromagnetic retarded interactions and (b) 1D model with long-range ferromagnetic retarded interactions }
	\label{fig:fig0}
\end{figure}

In this study, we address this issue by introducing interacting rotor models with time-delayed self-interactions driven by a memory-feedback mechanism. Specifically, we investigate a two-dimensional (2D) model with short-range antiferromagnetic(AFM) retarded interactions and a one-dimensional (1D) model with long-range ferromagnetic (FM) memory (Fig.\ref{fig:fig0}). In 2D, the system supports a novel class of topological defects—helical vortices with opposite vorticities propagating in opposite directions along the domain wall. In 1D, long-range temporal interaction circumvents the Mermin-Wagner theorem\cite{Mermin1966,Hohenberg1967} by establishing true long-range order and a finite-temperature phase transition. While previous approaches to realizing orders forbidden by Mermin-Wagner relied on driving the system out of equilibrium\cite{Vicsek1995,Nakano2021} or coupling it to an external bath\cite{Weber2022}, here we propose an alternative route by introducing memory effects into the many-body system.  Our work highlights engineered time-delay feedback as a versatile route to explore non-equilibrium physics.

{\it Model--} We consider two models of interacting classical rotors, where each rotor is coupled to an independent thermal bath and interacts with its nearest neighbors. Additionally, it is subjected to a memory force determined by its history. The equation of motion (EOM) for the $i\text{th}$ rotor reads
\begin{equation}I\ddot{\theta}_i(t)+\gamma \dot{\theta}_i(t)=F_i^s(t)+F_i^r(t)+\xi_i(t), \label{eq:EOM}\end{equation}
where $\theta_i(t)\in (-\pi,\pi]$ is the angle of rotor i and $I$ denotes its moment of inertia. The coupling to a thermal bath introduces both a friction coefficient $\gamma$ and a stochastic force $\xi_i(t)$. They satisfy the classical fluctuation-dissipation theorem $\langle \xi_i(t)\xi_j(t')\rangle_\xi=2k_B T \gamma \delta_{ij}\delta(t-t')$ where $T$ is the heat bath temperature and $\langle~ \rangle_\xi$ denotes an ensemble average over the noise trajectories. We assume that the rotors are coupled to their nearest neighbors via XY-type interactions of strength $J$, which exert a torque $F_i^s(t)$ on the $i\text{th}$ rotor given by $F_i^s(t)=-J\sum_{\delta}\sin[\theta_i(t)-\theta_{i+\delta}(t)]$
where the summation is over all the nearest neighboring sites of $i$ in the lattice.

\begin{figure*}[htbp]
	\centering
	\includegraphics[width=0.95\textwidth]{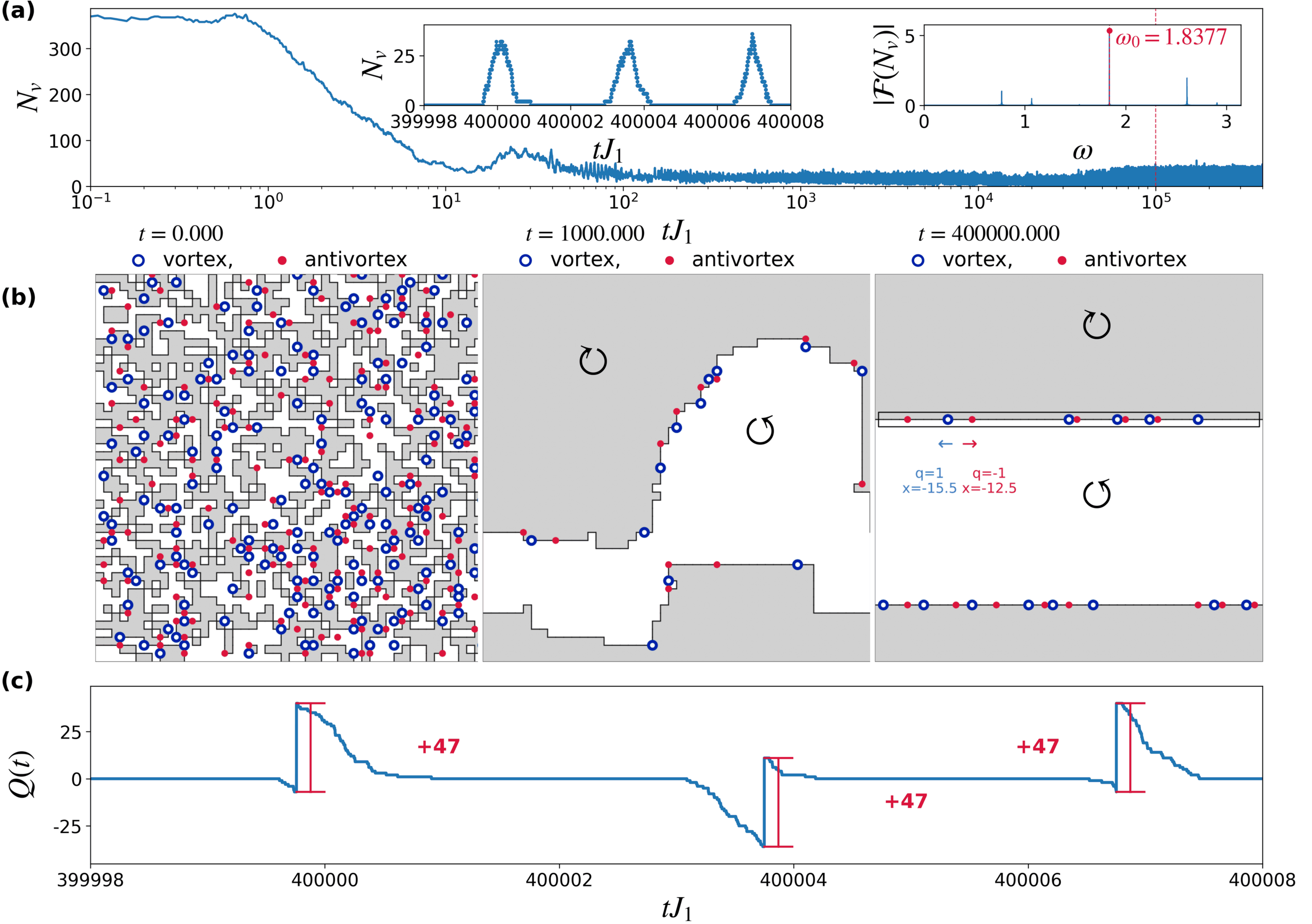}
	\caption{(a) Dynamics of the total vortex number $N_v$ following a quench from $T=J$ to $0.05J$. Left inset: late-time dynamics of $N_v$. Right inset: Fourier spectrum showing a peak at $\omega_0\approx 2\omega_c$. (b) Snapshots of DWs and vortices at initial, intermediate, and late times. Gray (white) regions indicate clockwise (counterclockwise) rotation; open (solid) circles denote vortices (antivortices). (c) The net dipole moment $Q(t)$ along the upper DW in (b). Parameters: $J_1=\gamma=I=J$, $\tau_c=2J^{-1}$. }
	\label{fig:fig1}
\end{figure*}

The key ingredient of our model is the time-delayed self-interaction $F_i^r(t)$. In its absence, the system relaxes to the thermodynamic equilibrium state of XY model with temperature T. This memory force represents a non-local history dependence, taking the form
\begin{equation}F_i^r(t)=\int_0^t dt' K(t-t') \sin[\theta_i(t)-\theta_i(t')], \label{eq:Fr}
\end{equation}
We will consider two models with different memory kernel $K(t-t')$: a 2D model with short-range memory, and a 1D model with long-range memory. In both cases, the periodical boundary condition is considered.

To characterize the system's ordering and fluctuations, we define the steady-state temporal and spatial correlation functions of the rotor magnetization $m_i(t) = \cos\theta_i(t)$ and  angular velocities  $v_i(t)=\dot{\theta}_i(t)$ as
\begin{small}
\begin{eqnarray}
D(t)&=&\frac 1L\sum_i \frac 1{T_0} \int_{t_{\mathrm{eq}}}^{t_{\mathrm{eq}}+T_0} dt_0 \cos[\theta_i(t_0)-\theta_i(t_0+t)] \label{eq:Dt}\\
G(r)&=&\frac 1L\sum_i \frac 1{T_0} \int_{t_{\mathrm{eq}}}^{t_{\mathrm{eq}}+T_0} dt_0 \cos[\theta_i(t_0)-\theta_{i+r}(t_0)] \label{eq:Gr}\\
G_v(r)&=&\frac 1L\sum_i \frac 1{T_0} \int_{t_{\mathrm{eq}}}^{t_{\mathrm{eq}}+T_0} dt_0 v_i(t_0) v_{i+r}(t_0) \label{eq:Gv}
\end{eqnarray}  
\end{small}
Here, $t_{\mathrm{eq}}$ is the equilibration time required for the system to reach the steady state, and $T_0$ denotes a sufficiently long time window over which the temporal average is performed to improve statistical sampling.

{\it 2D model with short-range antiferromagnetic retarded interactions: emergence of helical vortices--} The first model we consider is a 2D model on a $L\times L$ square lattice with periodic boundary condition and kernel function:
\begin{equation}K(t-t')=J_1\delta(t-t'-\tau_c) \label{eq:shortrange}
\end{equation}
where $\tau_c$ is a characteristic delay time, and $J_1>0$ denotes an AFM coupling between the state of the same rotor at two time slices separated by an interval $\tau_c$, implying that the rotor orientations at times $t$ and $t+\tau_c$ tend to be antiparallel, which, in turn, suggests a preference for rotation at a constant velocity. 

We first consider a noiseless, overdamped {\it single-rotor} model ($I=J=T=0$) governed by the EOM:
\begin{equation}
	\gamma \dot{\theta}(t)=J_1 \sin[\theta(t)-\theta(t-\tau_c)]. \label{eq:single1}
\end{equation}
To derive an analytical solution to Eq.~(\ref{eq:single1}), we adopt the ansatz of a constant angular velocity $\omega$, such that $\theta(t)=\theta_0+\omega t$. Substituting this ansatz into Eq.~(\ref{eq:single1}) yields a time-independent self-consistent equation for $\omega$: $
	\gamma\omega=J_1 \sin(\omega\tau_c)$, 
For $J_1\tau_c/\gamma < 1$, it possesses only the trivial static solution $\omega=0$. In contrast, for $J_1\tau_c/\gamma > 1$, the static state becomes unstable, and non-zero solutions $\omega = \pm \omega_c$ emerge in pairs,  corresponding to rotating states spontaneously breaking the continuous time-translation symmetry (CTTS) of the system. Concurrently, it also breaks a discrete $\mathbb{Z}_2$ chiral symmetry through the random selection of either clockwise ($\omega_c$) or counterclockwise ($-\omega_c$) rotation. With finite noise, both the CTTS and the $\mathbb{Z}_2$ chiral symmetry are restored\cite{Supplementary}.

In 2D, weak thermal fluctuations convert long-range time-crystalline order into quasi-long-range one (End matter). In contrast, $G_v(r)$ displays true long-range order driven by $\mathbb{Z}_2$ chiral symmetry breaking. Although the stability of these emergent dynamical orders does not go beyond the scope of the conventional framework, the nonequilibrium feature of our model could indeed lead to exotic topological defects absent in equilibrium. In general, a topological defect is governed by the symmetry and dimensionality of system\cite{Mermin1979,Thouless1998}. In 2D, $\mathbb{Z}_2$ and $U(1)$ symmetries respectively dictate the formation of line defects (domain walls, DWs) and point defects (vortices). Possessing both symmetries, our model supports both types of defects: DWs separating domains with opposite chirality, $\mathrm{sign}(v_i) = \pm 1$, and vortices with  nonzero winding number $q_i$ of $\theta_i$ around a plaquette. Crucially, their interplay  gives rise to intriguing  dynamics. To  demonstrate it, we quench the system from a high-T  to a low-T state and monitor the  total vortex number, $N_v = \sum_i \vert{}q_i\vert{}$. As shown in Fig.~\ref{fig:fig1} (a),  following an initial rapid decay from a disordered state  with randomly distributed DWs and vortices, the system enters a metastable state where two well-separated, nearly parallel straight DWs  trap virtually all remaining vortices[Fig.~\ref{fig:fig1} (b)], whose number ($N_v$) oscillates at a frequency  $2\omega_c$ ($\omega_c=0.94J$) over a duration scaling exponentially with DW separation.

Consequently, it hosts two distinct vortex species: bulk vortices randomly distributed in space as in the standard 2D XY model, and DW vortices localized exclusively along DWs and oscillate coherently in number. Similar DW-trapping vortex dynamics was studied in non-equilibrium XY models\cite{Rouzaire2021} and multi-band superconductors\cite{Zheng2026}, albeit via different mechanisms. Crucially, DW-vortices display helical dynamics, with opposite vorticities propagating in opposite directions along the DW, resembling the helical edge state in the quantum spin hall systems\cite{Wu2006}. This helical transport can be quantified via the net dipole moment along an $x$-parallel DW (say, the upper DW in Fig.\ref{fig:fig4} b): $Q(t) = \sum_i q_i x_i$, where $q_i$ is the vorticity and the $i$-th plaquette intersected by this DM and $x_i=i-\frac L2+\frac 12$ is its coordinate. As shown in Fig.~\ref{fig:fig1}(c), $Q(t)$ decreases monotonically via discrete step transitions corresponding to single-vortex motion, interrupted by sudden jumps of magnitude $L-1$ as vortices cross the periodic ``boundary" ({\it e.g.} when a vortex hops from site 1 to L, $x_i$ is suddenly changed by $L-1$). This result provides direct evidence of helicity. The coherent oscillation and helical dynamics of can be understood via a simple model, as shown in the End Matter.

%To understand DW vortex dynamics, we first analyze a single plaquette intersected by a DW [Fig.~\ref{fig:fig5}(b)], where vertices 1 and 2 belong to one domain and vertices 3 and 4 belong to another with opposite chirality. Assuming the angular velocities of the rotors are fixed: $\theta_{1,2}(t)=\theta_{1,2}-\omega t$ (clockwise) and $\theta_{3,4}(t)=\theta_{3,4}+\omega t$ (counterclockwise).  Fig.~\ref{fig:fig5}(b) shows that the plaquette vorticity $C_i$ periodically jumps  whenever a phase difference, $\theta_{1}(t)-\theta_{3}(t)$ or $\theta_{2}(t)-\theta_{4}(t)$, crosses $\pi$ (see Appendix.\ref{sec:toymodel}). This enables coherent vortex generation and annihilation, contrasting sharply with bulk vortices that require thermal activation to overcome a finite excitation gap. An extension to three plaquettes (Appendix.\ref{sec:toymodel}) similarly explains helical DW vortex dynamics.

\begin{figure}[htbp]
	\centering
	\includegraphics[width=0.49\textwidth]{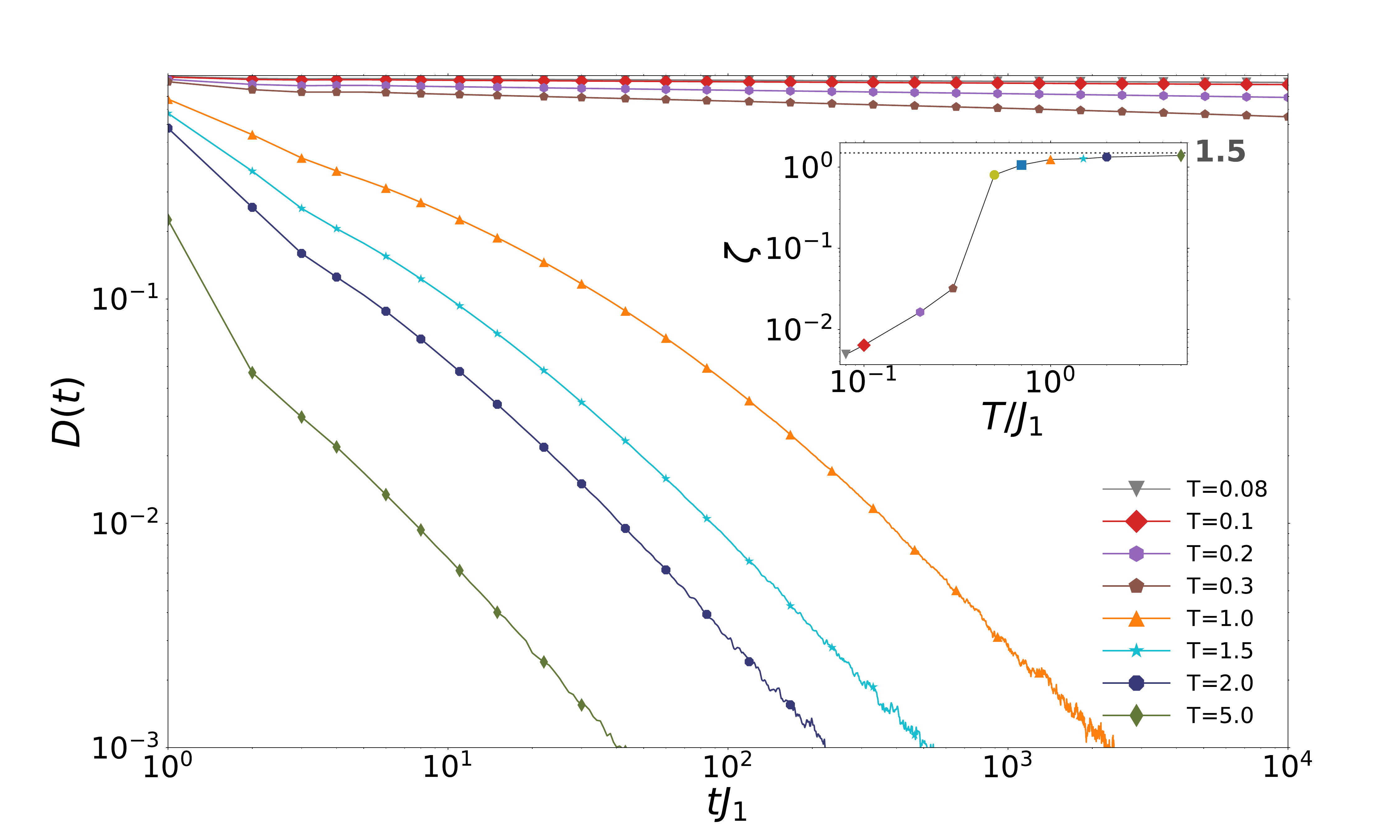}
	\caption{ $D(t)$ for a single rotor with $\alpha=3/2$ across various temperatures $T$, which exhibit power-law decay $D(t)\sim t^{-\zeta}$ in both the low- and high-T regimes. Inset: T-dependence of the exponent $\zeta$. Parameters: $I=\gamma=J_1$, $J=0$, $\tau_0=J^{-1}$.}
	\label{fig:fig2}
\end{figure}

\begin{figure*}[htb]
	\includegraphics[width=0.99\textwidth]{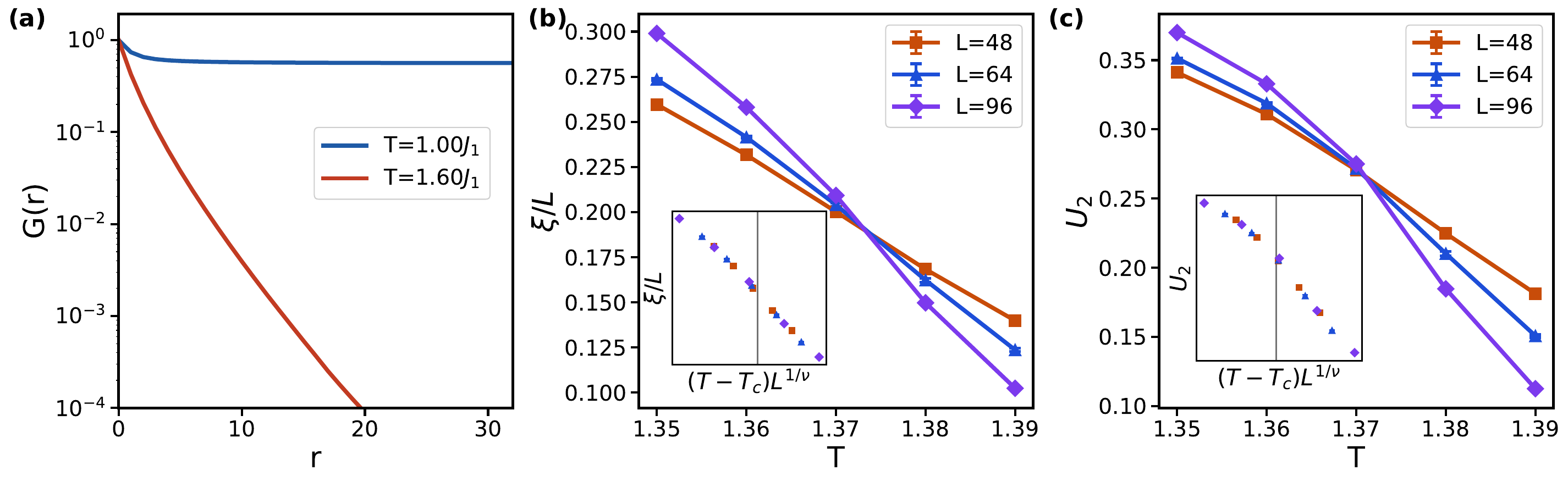}
	\caption{(a)$G(r)$ at low and high temperature; (b) Normalized correlation length $\xi/L$ and (c) Binder  cumulant  as a function of temperature for various $L$.  The insets are the data collapse with the optimal $\nu$.  Parameters: $I=\gamma=J_1=J$, $\tau_0=J^{-1}$.}
	\label{fig:fig3}
\end{figure*}

{\it 1D  model with long-range ferromagnetic retarded interactions--} The second model we consider is a 1D model  with long-range FM retarded interactions, whose memory kernel function in Eq.(\ref{eq:Fr}) is defined as:
\begin{equation}
	K(t-t') = -J_1(t-t'+\tau_0)^{-\alpha},	
\end{equation}
Here, $J_1 > 0$ is the FM retarded interaction strength, which decays algebraically over time with exponent $\alpha$. $\tau_0$ acts as a short-time cutoff regularizing the $t' \to t$ divergence. We fix $I=\gamma=J_1=\tau_0^{-1}=J$, leaving $\alpha$ and temperature $T$ as control parameters. We numerically solve Eq.~\eqref{eq:EOM} for its steady-state properties, where ensemble averages become time-independent. Reaching a steady state requires $\alpha > 1$; for $\alpha \le 1$, the divergence of the integral Eq.~\eqref{eq:Fr} over long time  yields aging or glassy dynamics rather than a true steady state.

%To track the system's dynamics, we solve the integro-differential Eq.~\eqref{eq:EOM} numerically. Because the memory kernel introduces a computational cost that scales quadratically with the simulation time $t$, we implement a truncation threshold $t_\mathrm{cut}$ such that $K(\tau)=0$ for $\tau>t_\mathrm{cut}$. The convergence of our numerical results with respect to $t_\mathrm{cut}$ has been verified (see the Supplementary Material [SM] for details). We focus on the long-time steady-state properties, where ensemble-averaged observables become time-independent. Attaining a well-defined steady state requires a sufficiently fast memory decay ($\alpha>1$); for $\alpha \le 1$, the integral of the kernel function $K(\tau)$ diverges over long times, yielding aging or glassy dynamics rather than a true steady state.

To illustrate the effect of long-range memory, we still focus on a single, overdamped rotor first. The resulting integro-differential EOM  reads
\begin{equation}
	\gamma \dot{\theta}(t)=\int_0^t dt' K(t-t') \sin[\theta(t)-\theta(t')] +\xi(t). \label{eq:EOM2}
\end{equation}
FM temporal coupling favors the alignment of the rotor with its historical states, yielding a static solution to Eq.~\eqref{eq:EOM2} if $\xi(t)=0$. To analyze the low-T regime, we linearize Eq.~\eqref{eq:EOM2}, $\sin[\theta(t)-\theta(t')]\approx \theta(t)-\theta(t')$ around the static state. Solving the linearized EOM via Fourier transform yields the frequency components\cite{Supplementary}.
\begin{equation}
	\theta(\omega)\approx \frac{\xi(\omega)}{i\gamma\omega-J_1\Gamma(1-\alpha)(i\omega)^{\alpha-1}}, \label{eq:Fourier1}
\end{equation}
where $\Gamma(1-\alpha)$ is the Gamma function, $\xi(\omega)$  satisfies $\langle \xi(\omega)\xi(-\omega)\rangle_\xi=2k_B T$. If $1<\alpha<2$, the memory term $|\omega|^{\alpha-1}$ dominates over the  viscous term $i\gamma\omega$ in the hydrodynamic limit ($\omega\rightarrow 0$), directly leads to a diverging power spectrum: $\langle\theta(\omega)\theta(-\omega)\rangle_\xi \sim T|\omega|^{2-2\alpha}$.

By performing the inverse Fourier transformation, we obtain the long-time scaling behavior of $D(t)$:
\begin{small}
\begin{eqnarray}
\nonumber &D(t)&=\langle \cos[\theta(t_0)-\theta(t_0+t)]\rangle=e^{-\frac 12 \langle [\theta(t_0)-\theta(t_0+t)]^2\rangle} \\
&=&e^{-\int_0^\Lambda d\omega  (1-\cos\omega t)\langle \theta(\omega)\theta(-\omega)\rangle}\sim  e^{-\int_{0}^{\Lambda} d\omega  \frac{T(1-\cos\omega t)}{|\omega|^{2\alpha-2}} } \label{eq:integral}
\end{eqnarray}
\end{small}
where $\Lambda \sim 2\pi/\tau_0$ is the ultraviolet cutoff regularized by the short-time cutoff $\tau_0$ of the algebraic memory kernel.

In the following, we focus on a marginal case $\alpha=3/2$, where the integral in the exponent of Eq.~\eqref{eq:integral}  diverges logarithmically as $T \ln (\Lambda t)$, yielding a power-law decay $D(t)\sim t^{-\zeta}$, with a T-dependent exponent $\zeta\propto T$. The linearization approximation holds in the weak-noise regime. To probe non-linear effects, we numerically simulate the EOM.~\eqref{eq:EOM} across various $T$, and find  $D(t)$ decays algebraically across all temperatures $D(t)\sim t^{-\zeta}$  (Fig.~\ref{fig:fig1}). At low $T$, $\zeta$ scales linearly with T (Fig.~\ref{fig:fig6}, inset), matching our analytic prediction. In contrast to the equilibrium 2D XY model whose correlations decay exponentially at high T\cite{Berezinskii1971,Kosterlitz1973}—our single-rotor model maintains algebraic decay at arbitrarily high $T$, with $\zeta$ monotonically approaching $\alpha$ as $T \rightarrow \infty$. This persistent power law  is reminiscent of the Griffiths inequalities for long-range interacting equilibrium systems whose correlation functions are bounded from below by the power law of the interaction potential itself\cite{Griffiths1969,Dunlop1977}.

 We now consider the 1D model, still focusing on $\alpha=3/2$ case.  To analyze it analytically, we first derive a quasi-continuous description, $\theta_i(t)\rightarrow \theta(x,t)$, based on a spin-wave approximation (SWA): $\sin[\theta_i(t)-\theta_{i+1}(t)]+\sin[\theta_i(t)-\theta_{i-1}(t)]\approx -\partial_x^2 \theta(x,t)$, and Eq.~\eqref{eq:EOM} reduces to
\begin{small}
\begin{equation}
	\gamma\partial_t\theta(x,t)=J\partial^2_x\theta(x,t)+\int_0^t dt' K(t-t') [\theta(x,t)-\theta(x,t')] +\xi(x,t).
\end{equation}
\end{small}
By performing a Fourier transform and taking the hydrodynamic limit $\omega\rightarrow 0$,$k\rightarrow 0$, we can obtain\cite{Supplementary}
\begin{equation}
	\langle \theta(k,\omega)\theta(-k,-\omega)\rangle\sim \frac{T}{c|\omega|+\sqrt{2c|\omega|} k^2+ k^4}. \label{eq:correlation4}
\end{equation}
where $c=\Gamma(-\frac 12)J_1$. Eq.~\eqref{eq:correlation4} demonstrates that the long-range memory effect qualitatively alters the dispersion relation, distinguishing it from a conventional FM spin wave. This leads to two key physical consequences. First,  Eq.~\eqref{eq:correlation4} yields the real-space scaling $\langle \theta(x,t)\theta(x+r,t)\rangle\sim |r|^{-1}$ for large $r$. Consequently,  $G(r)\sim e^{-a/r}$ approaches a constant at long distances, signaling true long-range order,  in stark contrast to the short-range correlations typical of conventional 1D FM systems. Second, the denominator of Eq.~\eqref{eq:correlation4} implies a dynamical critical exponent of $z=4$, differing significantly from the conventional FM value of $z=2$.

At low temperatures, SWA predicts true long-range order, which is confirmed numerically by Fig.~\ref{fig:fig3}(a), where $G(r)$ approaches a non-zero plateau at large distances for low $T$. At high $T$, $G(r)$ decays exponentially, marking the destruction of order by thermal noise.  This distinct behaviors of $G(r)$  signify a finite-temperature phase transition, whose critical properties can be extracted via the correlation length : $\xi = \frac{1}{q_0}\sqrt{\frac{S(0)}{S(q_0)}-1}$\cite{Sandvik2010},
 where $q_0 = 2\pi/L$ and $S(q) = \frac{1}{L} \sum_r e^{iqr} G(r)$ is the structure factor. The normalized correlation length $\tilde{\xi} = \xi/L$ as a function of $T$ for various  $L$ is plotted in Fig.~\ref{fig:fig3}(b). The curves exhibit a distinct crossing point at $T_c = 1.372J$, indicating a scale-invariant critical point. To determine the correlation length critical exponent $\nu$, we perform an optimal data collapse analysis, which yields $\nu = 1.18(2)$ [Fig.~\ref{fig:fig3}(b) inset]. $\nu$ can also be extracted from the Binder cumulant: $U_2=\frac 32 (1-\frac 13 \frac{\langle m^4\rangle}{\langle m^2\rangle^2})$\cite{Sandvik2010}, where $m=\frac 1L\sum_i\cos\theta_i$. Fig.~\ref{fig:fig3}(c) also exhibit a crossing point at $T=T_c$, and a data collapse analysis suggests $\nu=1.16(3)$.
We also examine $D(t)$ and $G(r)$ at the critical point. As shown in Fig.~\ref{fig:fig4}, both functions exhibit power-law decay: $D(t) \sim t^{-0.175}$ and $G(r) \sim r^{-0.686}$. By comparing their scaling exponents, we extract the dynamic critical exponent $z\approx 3.9(3)$, roughly agreeing with our analytical prediction $z=4$. 

\begin{figure}[htbp]
	\centering
	\includegraphics[width=0.49\textwidth]{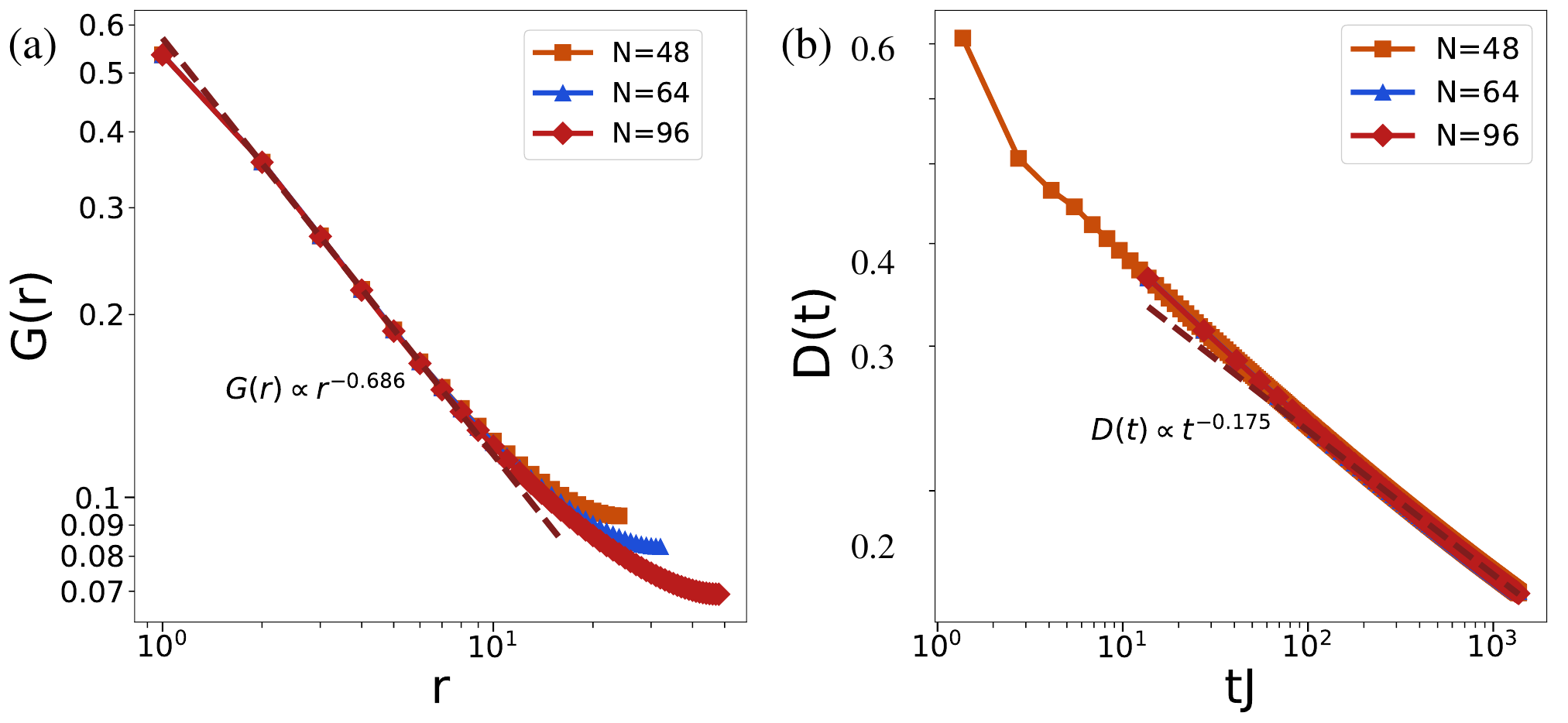}
	\caption{ (a) G(r) and (b) D(t) at critical point ($T_c=1.373J$) with various $L$. Parameters: $I=\gamma=J_1=J$,$\tau_0=J^{-1}$.}
	\label{fig:fig4}
\end{figure}

{\it Experimental realization--} To physically realize this model, we propose an active mechatronic metamaterial framework where spatial coupling, temporal memory, and thermal noise are digitally synthesized in real time\cite{Popa2013}. It has been widely used to explore the non-equilibrium physics \cite{Brandenbourger2019,Scheibner2020,Veenstra2024}.  The apparatus consists of a 1D or 2D array of  robotic nodes, each of which features a low-inertia  electric motor with torque actuation, a high-resolution  encoder to track the angular coordinate, and a current-controlled driver. A centralized real-time controller samples the angular states at a high frequency ($f_s \ge 1\text{ kHz}$). During each time step $\Delta t$, the controller records $\theta_i(t)$ into a local ring-buffer memory, computes the nearest-neighbor spatial torque $F_{i}^{s}(t)$ , and performs a discrete historical convolution for the retarded self-interaction $F_{i}^{r}(t) = \sum_{m} K(m\Delta t)\sin[\theta_{i}(t)-\theta_{i}(t-m\Delta t)]\Delta t$ based on the memory kernel. Finally, a digitally generated pseudo-random white-noise torque $\xi_i(t)$ calibrated to a software-defined temperature $T$ is injected to satisfy the classical fluctuation-dissipation theorem\cite{Ciliberto2017}. This synthesized net torque command is converted to a proportional motor current, allowing a setup to seamlessly tune parameters  to experimentally map out the predicted memory-driven phases.

{\it Conclusion and outlook--} In conclusion, we show that history-dependent temporal interactions induce nonequilibrium many-body physics with no equilibrium counterpart. Unlike conventional spatial interactions, temporal interactions are  inherently non-reciprocal due to strict causal constraints (a past state to influence the future, but not vice versa), drawing a  parallel to spatial non-reciprocal systems~\cite{Avni2023,Hanai2024,Belyansky2025,Avni2025,Lorenzana2025} (we contrast these systems with their one-dimensional higher spatial equilibrium analogs in End matter)f. Extending this classical framework to the quantum regime promises novel quantum dynamics. Additionally, it is highly desirable to identify natural open systems where the engineered memory kernel emerges intrinsically by integrating out environmental bath degrees of freedom\cite{Weiss1999,Leggett1987}, complementing our proposed engineered time-delayed feedback framework.
%and our work establishes a novel paradigm for exploring the unconventional non-equilibrium phases and phase transition based on active metamaterials capable of digitally processing temporal information. 

{\ Acknowledgments--} 
ZC is supported by the National Key Research and Development Program of China (2024YFA1408303), Natural Science Foundation of China (Grant No.12525407), Shanghai Science and Technology Innovation Action Plan(Grant No. 24Z510205936).

\bibliography{real}

\clearpage
\onecolumngrid 
\section*{End Matter}

\twocolumngrid

\begin{figure*}[htb]
	\includegraphics[width=0.99\textwidth]{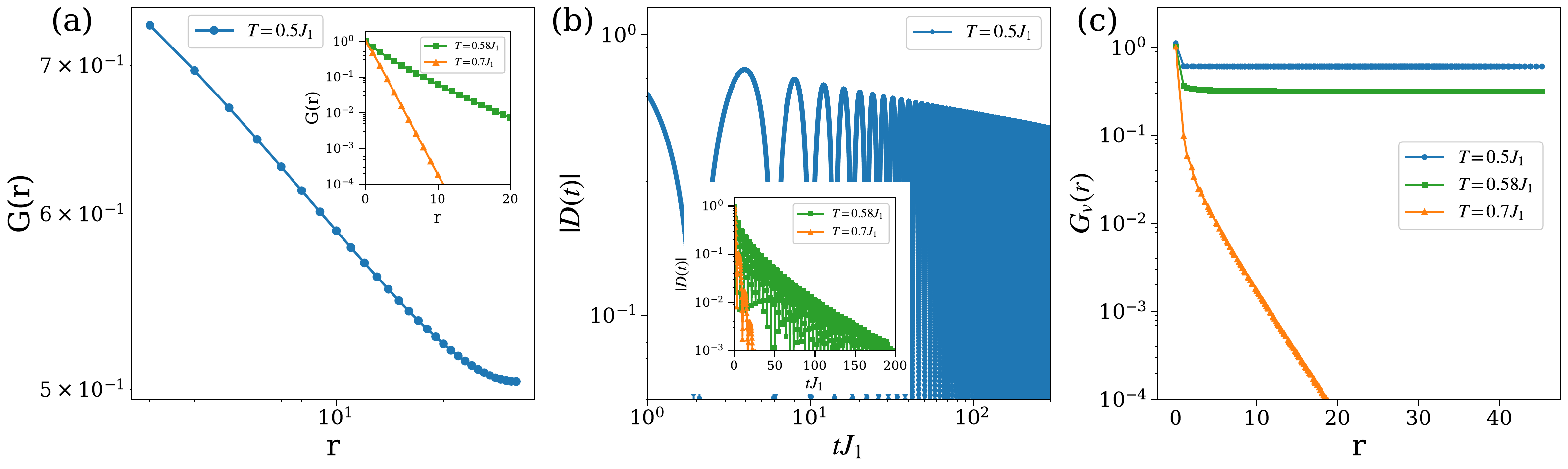}
	\caption{Spatial (a), temporal (b), and velocity (c) correlation functions at three representative temperatures across distinct phases. The log-log (semi-log) scale highlights algebraic (exponential) decay. Parameters: L=96, $I=\gamma=J_1=J$, $\tau_c=2J^{-1}$.}
	\label{fig:fig5}
\end{figure*}

\section{Stability of 2D non-equilibrium phases against thermal fluctuation}

As noted in the main text, a zero-temperature 2D rotor model with strong antiferromagnetic (AFM) short-range, time-delayed interactions eventually reaches a steady state where all rotors revolve synchronously with a uniform angular velocity. This state simultaneously breaks U(1) rotational symmetry ($\theta_i(t) \to \theta_i(t) + \theta_0, \forall i$), continuous time-translation symmetry ($\theta_i(t) \to \theta_i(t + t_0), \forall i$), and $\mathbb{Z}_2$ chiral symmetry ($v_i(t) \to -v_i(t), \forall i$). To evaluate the stability of these symmetry-broken phases against thermal fluctuations, we analyze the non-equilibrium phases of the 2D model by tracking correlation functions across three representative temperatures.
 
 As shown in Fig.~\ref{fig:fig5}, at low temperatures (e.g., $T=0.5J_1$), $G(r)$ decays algebraically with distance. $D(t)$ likewise exhibits algebraic decay superimposed with oscillations inherent to the single-rotor rotating solution, indicating that thermal fluctuations convert long-range time-crystalline order into quasi-long-range order. In contrast, $G_v(r)$ displays true long-range order driven by discrete ($\mathbb{Z}_2$) chiral symmetry breaking, which remains resilient against 2D thermal fluctuations. 
At intermediate temperatures (e.g., $T=0.58J_1$), thermal fluctuations eliminate quasi-long-range order of the magnetization—leading to exponentially decaying $G(r)$ and $D(t)$—while chiral long-range order persists.   Finally, at high temperatures (e.g., $T=0.7J_1$), thermal noise destroys all spatial and temporal order. 

Consequently, the system is expected to exhibit two phase transitions with increasing temperature. The first is defined by the concurrent breakdown of spatial and temporal quasi-long-range order, while the second corresponds to the restoration of $\mathbb{Z}_2$ chiral symmetry. Determining the universality classes of these transitions remains a subject for future study.

\begin{figure}[htbp]
	\centering
	\includegraphics[width=0.49\textwidth]{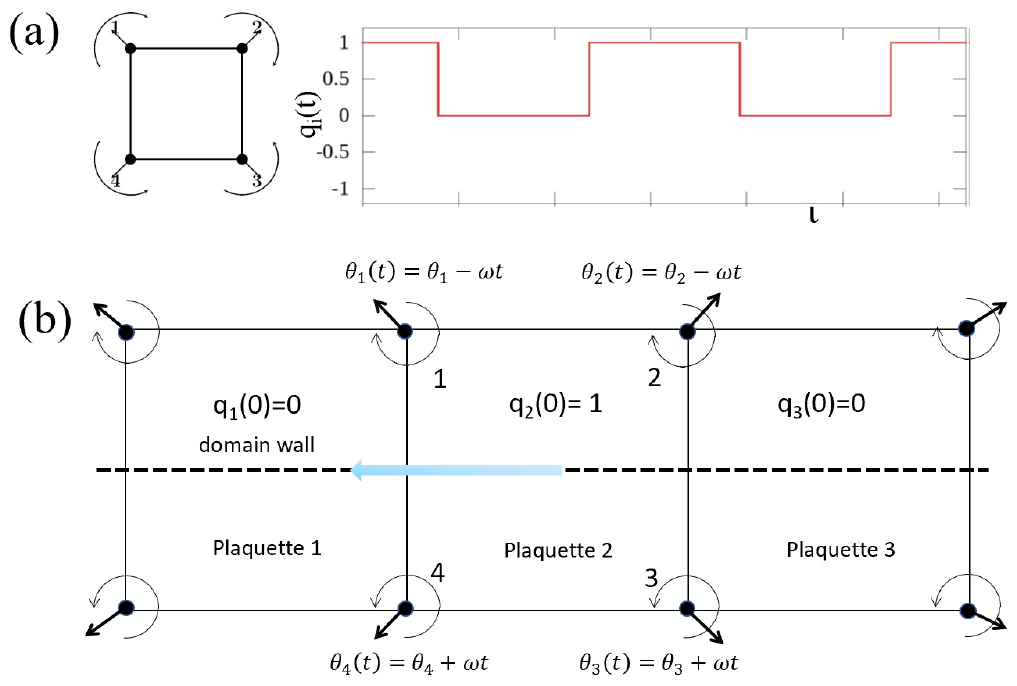}
	\caption{(a) Single-plaqutte toy model whose vorticity periodically oscillate in time and (b) Three-plaqutte toy model to illustrate the helical dynamics of a DW vortex.}
	\label{fig:fig6}
\end{figure}

\section{A toy model to illustrate the dynamics of doman-wall vortices}
The behavior of the domain-wall vortex, including its coherent oscillation and helical dynamics, can be understood via a simplified toy model. To illustrate the coherent oscillation of the velocity, we consider a single plaquette intersected by a DW (see Fig.~\ref{fig:fig6} (a), where vertices 1 and 2 belong to one domain and vertices 3 and 4 belong to another with opposite chirality. Assuming the angular velocities of the rotors are fixed as $\omega$ and rotors 1, 2 rotate clockwise and rotors 3, 4 rotate counterclockwise:
\begin{eqnarray}
	\nonumber \theta_1(t)&=&\theta_1-\omega t,~ \theta_2(t)=\theta_2-\omega t\\
	\theta_3(t)&=&\theta_3+\omega t,~ \theta_4(t)=\theta_4+\omega t\label{eq:SM1}
\end{eqnarray}
where $\theta_i$ is the initial phase of rotor i. Let $\theta_i(t)\in [0,2\pi)$,  the vorticity of this plaquette at time $t$ is defined as:
\begin{equation}
	q(t)=\frac{1}{2\pi}[\delta\theta_{14}(t)+\delta\theta_{43}(t)+\delta\theta_{32}(t)+\delta\theta_{21}(t)] \label{eq:vorticity}
\end{equation} 
where $\delta\theta_{ij}$ is the principal phase difference defined as:
\begin{equation}
	\delta\theta_{ij}(t)=\theta_j(t)-\theta_i(t)+2\pi m_{ij}(t)
\end{equation}
Here, $m_{ij}(t)$ is an integer chosen such that $\delta\theta_{ij}(t)\in (-\pi,\pi]$. Note that $m_{ij}(t)$ is generally time-dependent, changing by $\pm 1$ whenever $\theta_j(t)-\theta_i(t)$ crosses $\mp \pi$. Under the condition of Eq.~(\ref{eq:SM1}), $\delta\theta_{12}(t)$ and $\delta\theta_{34}(t)$ are stationary; thus, $q(t)$ changes only when $\theta_4(t)-\theta_1(t)=\theta_4-\theta_1+2\omega t$ or $\theta_2(t)-\theta_3(t)=\theta_2-\theta_3-2\omega t$ crosses $\pm\pi$. One can also prove that $q(t)$ jumps periodically between 0 and $\pm 1$, with a period equal to half that of the rotors.

The helical dynamics of the vortex can also be understood using a similar model consisting of three plaquettes intersected by a DW, where the upper and lower rotors rotate clockwise and counterclockwise, respectively, with identical angular velocities $\omega$ [Fig.~\ref{fig:fig6}(b)]. Assume that initially a vortex with vorticity $q_2(0)=1$ is located on the central plaquette, while the remaining two plaquettes contain no vortex, $q_1(0)=q_3(0)=0$. As established above, when $\theta_4(t)-\theta_1(t)$ crosses $\pm\pi$, $q_2(t)$ and $q_1(t)$ change value simultaneously, corresponding to a leftward shift of the vortex. Similarly, if $\theta_2(t)-\theta_3(t)$ crosses $\pm\pi$ first, $q_2(t)$ and $q_3(t)$ change simultaneously, resulting in rightward motion.

Next, we prove that $\theta_4(t)-\theta_1(t)$ always crosses $\pm\pi$ before   $\theta_2(t)-\theta_3(t)$ under the initial condition $q_2(0)=1$ and $q_1(0)=q_3(0)=0$. The dynamics can be categorized into four cases based on the principal phase differences $\delta\theta_{14}(0)$ and $\delta\theta_{32}(0)$ in plaquette 2 at $t=0$:

\begin{enumerate}
	\item If $-\pi \le \delta\theta_{14}(0),\delta\theta_{32}(0)<0$, according to Eq.~(\ref{eq:vorticity}), $q_2(0)\le 0$, contradicting the initial condition.
	
	\item If  $0\le \delta\theta_{14}(0),\delta\theta_{32}(0)<\pi$ ,  at short times $\delta\theta_{14}(t)=\delta\theta_{14}(0)+2\omega t$ and $\delta\theta_{32}(t)=\delta\theta_{32}(0)-2\omega t$, $\delta\theta_{14}(t)$ crosses $\pi$ prior to $\delta\theta_{32}(t)$.
	
	\item If $0\le \delta\theta_{14}(0)<\pi$ and $-\pi\le\delta\theta_{32}(0)<0$, then $\vert{}\delta\theta_{14}(0)\vert{}>\vert{}\delta\theta_{32}(0)\vert{}$  (otherwise, $q_2(0)< 1$, contradicting the initial state). Consequently, $\delta\theta_{14}(t)=\delta\theta_{14}(0)+2\omega t$ crosses $\pi$ before $\delta\theta_{32}(t)=\delta\theta_{32}(0)-2\omega t$ crosses $-\pi$.
	
	\item Similarly, if $-\pi\le \delta\theta_{14}(0)<0$ and $0\le\delta\theta_{32}(0)<\pi$, then $\vert{}\delta\theta_{14}(0)\vert{}<\vert{}\delta\theta_{32}(0)\vert{}$ to ensure $q_2(0)=1$. Thus, $\delta\theta_{14}(t)=\delta\theta_{14}(0)+2\omega t$ again crosses $\pi$ before $\delta\theta_{32}(t)=\delta\theta_{32}(0)-2\omega t$ crosses $-\pi$.	
\end{enumerate}
Therefore, given $q_2(0)=1$ and $q_1(0)=q_3(0)=0$, a vortex always moves leftward. Analogously, an antivortex  can be shown to move rightward, establishing the helical dynamics of the vortex.

\section{Non-equilibrium systems with temporal interaction  v.s. Equilibrium systems with one dimension higher}

To highlight the role of temporal interactions, we contrast our nonequilibrium systems with their spatial equilibrium one-dimensional higher analogues. Replacing the temporal axis in our first model with a spatial $z$-axis ($\tau_c$ serves as a lattice constant) yields a 3D anisotropic XY model with FM (AFM) coupling in the $xy$-plane (along $z$). This 3D equilibrium model exhibits true long-range order at low temperatures, contrasting sharply with the quasi-long-range order in our 2+1D model. 

The spatial analogue of our second model is a 2D anisotropic XY model featuring power-law interactions along the vertical axis with decay exponent $\alpha = 3/2$:
\begin{small}
	\begin{equation}
		H=\sum_{i_x,i_y} \big[-J\cos(\theta_{i_x,i_y}-\theta_{i_x+1,i_y}) -J\sum_r\frac{\cos(\theta_{i_x,i_y}-\theta_{i_x,i_y+r})}{r^{3/2}} \big] \label{eq:Ham3}
	\end{equation}
\end{small}
Under the spin-wave approximation, in long-wave limit, the Hamiltonian in Eq.(\ref{eq:Ham3}) can be expressed  as:  
\begin{equation}
	H=\int dk_x dk_y [k_x^2+|k_y|^{\frac 12}]\theta(\mathbf{k})\theta(-\mathbf{k}) \label{eq:Ham4}
\end{equation}

A key quantity to characterize this model with long-range interaction is the effective dimension $d_{\mathrm{eff}}$, which is  used to describe the dimensionality that a standard, isotropic, short-range model would need to  share the same phase space volume constraints and fluctuation behavior at the critical point. $d_{\mathrm{eff}}$ is determined by how the momentum-space volume element $d^2k = dk_x \, dk_y$ transforms under system rescaling. For Hamiltonian~(\ref{eq:Ham4}) to remain invariant under a uniform rescaling of the energy/temperature terms, the anisotropic momentum components must obey $\vert{}k_y\vert{}^{1/2} \sim k_x^2$. Scaling the horizontal momentum as $k_x \to b^{-1} k_x$ requires $k_y \to b^{-4} k_y$, so the volume element transforms as:$dk_x \, dk_y \to b^{-5} dk_x \, dk_y$. In an isotropic $d_{\mathrm{eff}}$-dimensional system, every direction scales as $b^{-1}$, yielding a volume factor of $b^{-d_{\mathrm{eff}}}$. Equating the scaling exponents gives $d_{\mathrm{eff}} = 5$.

Although Hamiltonian~(\ref{eq:Ham3}) is defined on a flat 2D lattice, the strong vertical long-range coupling ($\alpha = 3/2$) heavily suppresses $y$-directional fluctuations, expanding the effective phase space to $d_{\mathrm{eff}} = 5$. Since $d_{\mathrm{eff}} > 4$ exceeds the upper critical dimension, critical fluctuations are quenched, and the phase transition is governed by Landau mean-field theory. While the corresponding 2D equilibrium system also exhibits a long-range ordered phase, spin-wave analysis shows its transition belongs to the Gaussian universality class, with critical exponents ($\nu = 0.5$, $z = 2$) distinct from our non-equilibrium values ($\nu \approx 1.18$, $z = 4$).

These comparisons show that despite the similarities, the fundamental difference between spatial and temporal dimensions yields profound physical consequences that distinguish our non-equilibrium framework from its equilibrium counterparts.

\end{document}